\documentclass[doc]{article}
\usepackage[utf8]{inputenc}
\usepackage{amsmath}
\usepackage[a4paper, margin=1in]{geometry}
\usepackage{authblk}
\usepackage{graphicx}
\graphicspath{ {./images/} }
\usepackage{caption}
\usepackage{subcaption}
\usepackage[table,xcdraw]{xcolor}
\usepackage[parfill]{parskip} 
\usepackage{apacite}
\usepackage{siunitx}
\usepackage{float}
\usepackage{placeins}
\title{Beyond Suboptimality: Resource-Rationality and Task Demands Shape the Complexity of Perceptual Representations}
\author[1,4,*]{Andrew Jun Lee}
\author[3]{Daniel Turek}
\author[2,4]{\"{O}. Da\u{g}lar Tanr{\i}kulu}
\affil[1]{Department of Psychology, University of California, Los Angeles}
\affil[2]{Department of Psychology, University of New Hampshire}
\affil[3]{Department of Mathematical Sciences, Lafayette College}
\affil[4]{Department of Psychology, Williams College}
\affil[*]{Corresponding Author: andrewlee0@ucla.edu}

\begin{document}
\date{}
\maketitle


\begin{abstract}
\noindent
Early theories of perception as probabilistic inference propose that uncertainty about the interpretation of sensory input is represented as a probability distribution over many interpretations---a relatively complex representation. However, critics argue that persistent demonstrations of suboptimal perceptual decision-making indicate limits in representational complexity. We contend that suboptimality arises not from genuine limits, but participants' resource-rational adaptations to task demands. For example, when tasks are solvable with minimal attention to stimuli, participants may neglect information needed for complex representations, relying instead on simpler ones that engender suboptimality. Across three experiments, we progressively reduced the efficacy of resource-rational strategies on a carefully controlled decision task. Model fits favored simple representations when resource-rational strategies were effective, and favored complex representations when ineffective, suggesting that perceptual representations can be simple or complex depending on task demands. We conclude that resource-rationality is an epistemic constraint for experimental design and essential to a complete theory of perception.
\end{abstract}

\section{Introduction}
\begin{figure}[!b]
\centering
\includegraphics[width=0.75\textwidth]{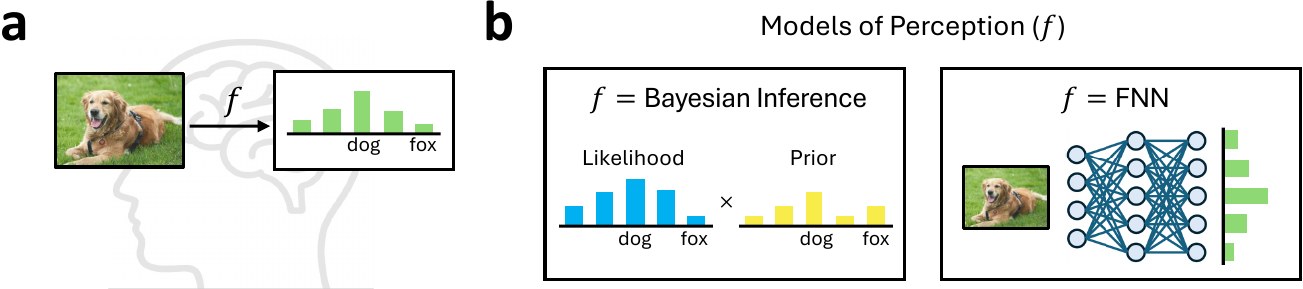}
\caption{Perception as probabilistic inference. (a) Perception is often modeled as a function $f$ that performs probabilistic inference, evaluating the plausibility of many possible hypotheses (e.g., plausibility of ``dog" versus ``fox" as the object's class). (b) Such inference has been implemented as Bayesian inference and feedforward neural networks (FNNs).}
\label{fig:intro_figureA}
\end{figure}

In recent decades, the view of perception as a form of probabilistic inference has become the dominant framework of perceptual psychology (\shortciteNP{fiser2010statistically, haefner2016perceptual, kersten2003bayesian, knill2004bayesian, ma2023bayesian, mamassian2002bayesian, peters2024does, pouget2013probabilistic, tanrikulu2021kind, zemel1998probabilistic}). According to this view, perception addresses the inherent ambiguity of sensory input by evaluating the plausibility of multiple interpretations or \textit{hypotheses} (Figure~\ref{fig:intro_figureA}a). This has been traditionally formalized in models of Bayesian inference, which specify how prior knowledge should update with sensory input, and more recently (and more contentiously) as feedforward neural networks (Figure~\ref{fig:intro_figureA}b; \shortciteNP{kriegeskorte2015deep, kietzmann2017deep, wichmann2023deep}). Growing behavioral and neural evidence for Bayesian accounts and neurons approximating probability distributions led to the Bayesian coding hypothesis (BCH), which states that perception represents interpretations of sensory input probabilistically, specifically in the form of probability distributions \shortcite{fiser2010statistically, knill2004bayesian, ma2006bayesian, pouget2013probabilistic, zemel1998probabilistic}. 

However, BCH has been highly contested. Researchers have proposed alternative representations, such as iteratively-formed samples of distributions \shortcite{sanborn2016bayesian}, summary statistics (e.g., mean and variance; \shortciteNP{rahnev2017case}), and truncated probability distributions (e.g., a distribution containing just the most probable hypotheses; \shortciteNP{rahnev2017case, rahnev2021perception}). These representations, critics claim, can explain evidence for BCH in canonical tasks, including two-alternative forced choice and Gaussian multisensory cue combination. Critics also state that (1) persistent demonstrations of suboptimal decision-making is evidence against Bayesian processes and probabilistic representations \shortcite{rahnev2018suboptimality, gardner2019optimality,yr}, and (2) purported direct evidence of probabilistic representations is confounded because it arises after multiple stimulus presentations  \shortcite{chetverikov2019feature, knill2007learning, kording2004bayesian, tanrikulu2020encoding, tanrikulu2021kind} or researchers mistake participants' sensitivity to uncertainty (e.g., varied behavioral responses) as a representation of uncertainty (\shortciteNP{block2018if, rahnev2021perception}; c.f., \shortciteNP{jabar2022perception}). Unfortunately, progress has recently stagnated due to a lack of consensus on the definition of probabilistic representations \shortcite{rahnev2021perception, rahnev2022mystery}. It is unclear whether probabilistic representations must conform to Kolmogorov probability axioms (a \textit{strong} definition; \shortciteNP{rahnev2021perception}), such that samples and truncated distributions are non-probabilistic; or merely encode some ``sense" of uncertainty (e.g., not a ``point estimate"), such as a large sample which, despite lacking explicit probabilities, implicitly encodes distributional information \shortcite{sanborn2016bayesian}. 

\begin{figure}[!b]
\centering
\includegraphics[width=\textwidth]{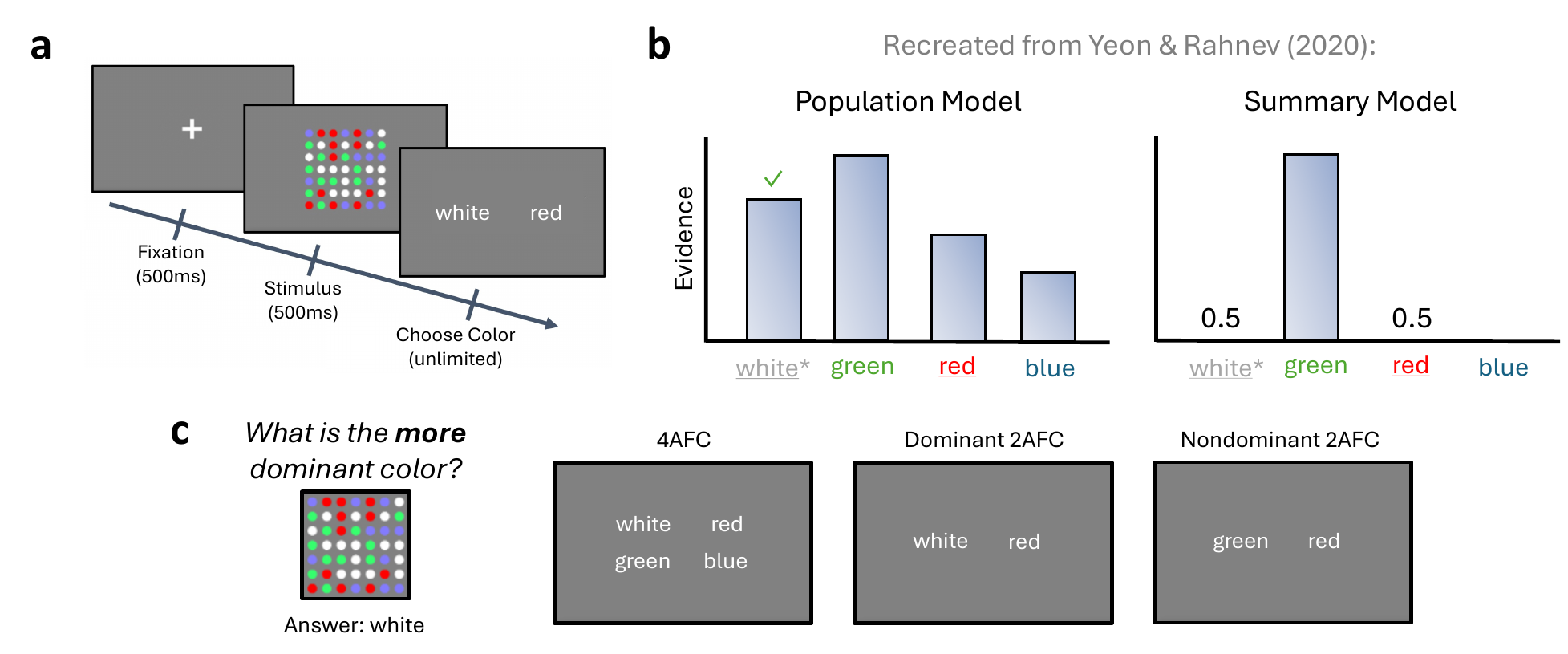}
\caption{Methodology of the current study based on Yeon and Rahnev (2020). (a) On each trial of their task, participants briefly viewed a fixation cross followed by an array of colored circles and reported the dominant color from a few options. (b) Figure recreated from Yeon and Rahnev (2020). This task can be solved by various decision models. The population model represents the full distribution of colors, the summary model represents only the perceived dominant color, and the two-highest model represents the two perceived most dominant colors. In (a), white is the dominant color. As Yeon and Rahnev (2020) explain, if the dominant color is misrepresented (e.g., ``green” instead of white), model decisions diverge: the population model can still select white by using information about other alternatives, whereas the summary model must guess because ``green” is not among the 2AFC options. (c) Trial types used in the current study based on response options. In ``4AFC" and ``dominant 2AFC trials," the dominant color is always among the options, so correct responses can be made without maintaining a full distribution. Our design adds nondominant 2AFC trials, where the dominant color is excluded and participants must choose the more frequent of the remaining options. These trials require encoding the full distribution, since they can only be solved if multiple alternatives are represented at the decision stage.}
\label{fig:intro_figureB}
\end{figure}

To make progress, we make two suggestions. First, we agree with \shortciteA{rahnev2022mystery} that researchers should distinguish the \textit{complexity} of perceptual representations from their exact format (e.g., samples, distributions, etc.). This allows us to avoid subjective debate over definitions that reflect researchers' preferences \shortcite{rahnev2022mystery} and to return to the central question, which we frame in terms of complexity: to what extent are perceptual representations simple (e.g., small samples \textit{\textbf{or}} truncated distributions) versus complex (e.g., large samples \textit{\textbf{or}} full distributions)? Second, any task claiming evidence for complex representations (which have been, most frequently, for full distributions) should avoid aforementioned critiques by, for example, probing information after one stimulus exposure, presenting more than two response options (in forced-choice designs), and explicitly comparing alternative representations. For example, \shortciteA{yr} used a multi-alternative forced-choice task in which participants briefly viewed an array of colored circles and selected the dominant color (the most frequent) from two or all four possible colors (Figure~\ref{fig:intro_figureB}a). Decision models with varying representational complexity about which color is dominant were fit to the response data (Figure~\ref{fig:intro_figureB}b). The \textit{population model} selected the most probable dominant color from a distribution over all four colors, the most complex representation. The \textit{two-highest model} selected from the two most probable colors of that distribution, a simpler representation. The \textit{summary model} selected from the single most probable color, the most simple representation. And the \textit{two-} and \textit{three-attention models} selected from two and three random colors. Model fits consistently favored the summary model, suggesting that representations are highly simple (e.g., a single sample or truncated distribution of just the maximum \textit{a posteriori}).

\begin{figure}[!b]
\centering
\includegraphics[width=0.75\textwidth]{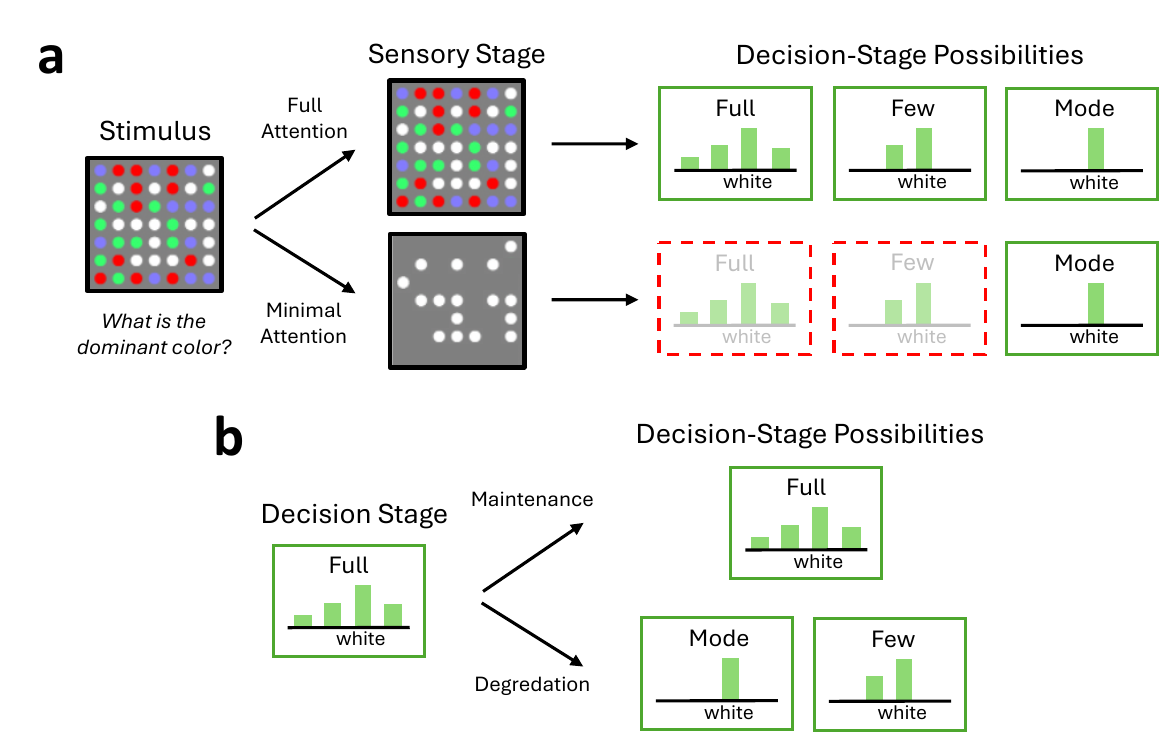}
\caption{When selecting the dominant color in an array of circles, participants can adopt at least two resource-rational strategies. (a) If they attend to the full stimulus (top row), decision-stage representations could contain the entire distribution or only a subset of alternatives. If they attend only to the dominant color (bottom row), sensory representations are impoverished, forcing simple decision-stage representations. In this case, more complex representations are unavailable (red borders). (b) Even with full attention, complex representations can degrade during the trial if not actively maintained.}
\label{fig:intro_figureC}
\end{figure}

However, conclusions denying complex representations are warranted only when task success requires using them. If success is possible without information about many or all hypotheses, participants may solve tasks using resource-rational strategies that result in simple representations, even if more complex representations can be formed \shortcite{lieder2020resource, lee2023human, lee2025enhancing, corral2026encouraging}. How might such simplification occur? \shortciteA{yr} distinguished between \textit{sensory representations} (early perceptual encodings of a stimulus) and \textit{decision-stage representations} (about which the debate on probabilistic representations is concerned) that are formed from sensory representations. We contend that resource-rational strategies can bottleneck information at either stage: if only a subset of stimulus information is useful for a task, participants may reduce attention to other features, leading to impoverished sensory representations and thus simple decision-stage representations (Figure~\ref{fig:intro_figureC}a); if information in decision-stage representations is not actively retained after forming, potentially extant distributional information will degrade (Figure~\ref{fig:intro_figureC}b). This concern is illustrated in the task of \shortciteA{yr}: since the dominant color is always a response option, participants can succeed by focusing exclusively on identifying the dominant color and ignoring or discarding information about others---a strategy that inadvertently produces simple representations and apparent suboptimal decisions.

The present study reworks this task to test whether the seemingly suboptimal performance reported in previous studies reflects a true limitation of perceptual decision-making circuits or instead arises from incomplete encoding at the sensory stage or degraded information at the decision stage, caused by demands imposed by the task. Across three experiments, task success increasingly required the use of distributional information and reduced the efficacy of resource-rational strategies. We compared only the population, two-highest, and summary models. Model fits shifted accordingly: when task demands were low, behavior was best explained by the summary model (Experiment 1); as the task changed, by the two-highest model (Experiment 2); and under the most demanding conditions, by the population model (Experiment 3).

\section{Experiment 1}
Experiment 1 examines whether previously observed suboptimal perceptual decision-making reflects limited processing capacity at the decision stage or representational simplification engendered by resource-rational adaptations to task demands. Building on the task of \shortciteA{yr}, we added two-alternative forced choice (2AFC) trials without the most dominant color. We refer to these trials as \textit{nondominant 2AFC} and trials including the dominant color as \textit{dominant 2AFC}. In all trials, participants were instructed to select the \textit{more} dominant color, not \textit{the} dominant color. Each participant completed two phases: a \emph{standard} experimental phase mirroring the original design, and an \emph{adjusted} version of the standard phase where half of the dominant 2AFC trials in the standard phase were replaced with nondominant 2AFC trials. All trials were interleaved in both phases. We predicted that the adjusted phase would yield stronger evidence for the summary model and weaker evidence for alternative models compared to the standard phase, since success on nondominant 2AFC trials, appearing unpredictably throughout the adjusted phase, requires representing the full distribution of colors.

\subsection{Method}
\subsubsection{Participants}
Thirty undergraduates from Williams College ($n = 18$) and the University of New Hampshire (UNH; $n = 12$) participated for course credit and provided informed consent. All participants had normal or corrected-to-normal visual acuity. Sample size was based on Yeon and Rahnev (2020) as an informed minimum.

\subsubsection{Stimuli}
The stimulus of all trials was a 7-by-7 array of circles (in the center of the screen) on a mid-gray background, in which each color was randomly colored either white, green, red, or blue. On each trial, one of the four colors was dominant, appearing in 16 circles, while the other three were nondominant, each appearing in 11 circles (Figure~\ref{fig:color_distribution}, left panel). The color of each circle was chosen randomly to ensure an even number of presentations for each color. The diameter of each circle was 1.57 degrees of visual angle (40 pixels), with 2.35 degrees (60 pixels) between the centers of adjacent circles. The total area of the array was 400 pixels. The experiment was completed in a dim room, participants sitting 57 cm away from the screen using a chin-rest.

\subsubsection{Procedure}
Each trial began with a fixation cross (500ms), followed by the stimulus (500ms), and ended with either a 4AFC or 2AFC response without time pressure (Figure~\ref{fig:intro_figureB}a). The task was to select the \textit{more} dominant color from the response options (Figure~\ref{fig:intro_figureB}c). Accuracy feedback was not provided. Participants completed three types of trials. In 4AFC trials, participants were presented with all four color options. In dominant 2AFC trials, participants chose between two options: one was always the dominant color, and the other was a randomly selected nondominant color. In nondominant 2AFC trials, both options were randomly selected nondominant colors. Since each nondominant color appeared in the stimulus an equal number of times (11 each), the nondominant 2AFC trials had no objectively correct answer. The nondominant 2AFC trials were crucial because success on them prevents participants from focusing solely on the dominant color and requires information about all colors.

The experiment consisted of two phases. In the standard phase, we included 4AFC and dominant 2AFC trials in a fixed ratio of one-third to two-thirds, to maintain comparability with Yeon and Rahnev's (2020) design. In the adjusted phase, all three trial types occurred in equal proportion as a consequence of replacing half of the dominant 2AFC trials with nondominant 2AFC trials. The adjusted phase aimed to examine the inclusion of nondominant 2AFC trials on model fits compared to the standard phase. We predicted that the adjusted phase would force participants to distribute their attention more evenly across colors or maintain decision-stage information about all colors, since the dominant color is no longer always the correct answer. In both phases, trial order was randomized, preventing the prediction of the next trial type, and the 4AFC-to-2AFC trial ratio was consistent (both two-thirds). Each phase consisted of 17 blocks of 72 trials (including 1 practice block), resulting in 1152 trials in total, excluding practice. Participants were unaware of the distinction between dominant and nondominant 2AFC trials, and practice blocks were excluded from analysis. Phase order was counterbalanced between-subjects. 

For the 18 Williams College participants, a one-minute break was given every 8 blocks starting from the first non-practice block. To increase engagement with the task, the 12 UNH participants were given 15-second breaks after each non-practice block and a point system awarding 20 points per correct answer (up to 720 points per block). While feedback was not provided during trials, cumulative scores (``Your current score is $n$/720'') and highest scores (``Your highest score is $n$/720'') were displayed during the 15-second break after each block. For non-dominant 2AFC trials, where no true answer existed in this experiment, a ``correct'' answer was randomly assigned for scoring purposes. Analyses confirmed no significant accuracy or reaction time differences between groups (see Supplemental Information).

\subsubsection{Modeling Framework}
We fit the decision models to human data following a procedure similar to that of \shortciteA{yr}, but used Bayesian inference instead of simulated annealing. To fit the decision models, we first modeled each participant's sensory representation as a single parameter, $\mu_p$, the mean evidence of the dominant color in phase $p$. (Evidence refers to a generic probability-like degree of belief.) Evidence for each of the three nondominant colors was modeled as having a mean of 0, since they have equal frequencies in the stimulus (not true in later experiments). The evidence values of a specific trial for the dominant and nondominant colors, $e_D$ and $e_{N1:3}$, were assumed to follow normal distributions with means of $\mu_p$ and 0 and standard deviations of 1: $e_D \sim \text{Normal}(\mu_p, 1)$ and $e_{N1}, e_{N2}, e_{N3} \sim \text{Normal}(0,1)$. Samples of these four evidences at trial \textit{i} simulate the sensory representation of trial $i$.

To estimate $\mu_p$, we used Bayesian inference on each participant's 4AFC accuracy data in phase $p$. Formally, the posterior over $\mu_p$ is the product of the prior over $\mu_p$ and the likelihood of an accurate trial $y_i=1$ conditioned on $\mu$ and trial $i$: $P(\mu_p|y=1)  \propto P(y_i=1|\mu_p) \cdot P(\mu_p)$. We used a flat, uninformative normal prior $P(\mu_p) = \mathcal{N}(mean=0, sd=5)$ and set the likelihood function to the indicator function $P(y_i=1|\mu_p) = 1$ if $e_D > \max(e_{N1}, e_{N2}, e_{N3})$ and $P(y_i=1|\mu_p)=0$ otherwise, where a trial $i$ is correct if the sampled evidence of the dominant color $e_D$ is greater than the sampled evidences of the nondominant colors $e_{N1:3}$. The posterior was estimated as a set of 500,000 samples of $\mu_p$ using slice sampling in the NIMBLE package in R \shortcite{nimble, neal2003slice}. 

After estimating the posterior for each phase of each participant, we evaluated model fit on dominant 2AFC accuracy data for a phase $p$. We used the mean of the posterior samples $\bar{\mu}_p$ as the point estimate for $\mu_p$ to compute Akaike Information Criterion (AIC), a measure of model fit (Supplemental Information). On each dominant 2AFC trial, we generated a sensory representation $\{ e_D,e_{N1:3}\}$ and applied the decision rules of each model, considering all four samples, the two highest, or the highest for the population, two-highest, and summary models, respectively. When a model did not have enough information, it guessed randomly (only non-population models).

Importantly, we did not fit the models directly to 2AFC data. On 2AFC trials, each model applies different decision-stage processes by filtering the sensory representation in different ways. If we estimated $\mu_p$ directly from 2AFC trials, the estimate would differ based on whichever decision model is used for the likelihood function. This would introduce a circularity in model fitting by using each model to estimate $\mu$, and then reusing that model-specific $\mu$ to evaluate the model's fit. Such circularity would make model comparisons invalid. Instead, we estimated $\mu_p$ from 4AFC trials, on which all models use the same decision rule and therefore predict the same response \shortcite{yr}. This ensures that $\mu_p$ is independent of the decision model and the same estimate is used when comparing models on 2AFC trials. More details about the modeling can be found in the Supplemental Information. 

Bayesian inference offers several advantages over other optimization methods. Slice sampling thoroughly explores the posterior, which reduces risk of converging to local optima. The estimated posterior can be assessed with rigorous diagnostic checks, including effective sample size, Gelman-Rubin convergence, and posterior predictive checks (Supplemental Information). We increased the 500,000 sample size of chains whenever diagnostics were not met.

\subsection{Results}
Trial data for each participant was excluded if reaction time fell below 50 ms or exceeded 4 standard deviations from their mean reaction time. On average, 11.83 trials (1.03\%) were dropped from the 1152 experimental trials. 

\begin{figure}[!b]
\centering
\includegraphics[width=\textwidth]{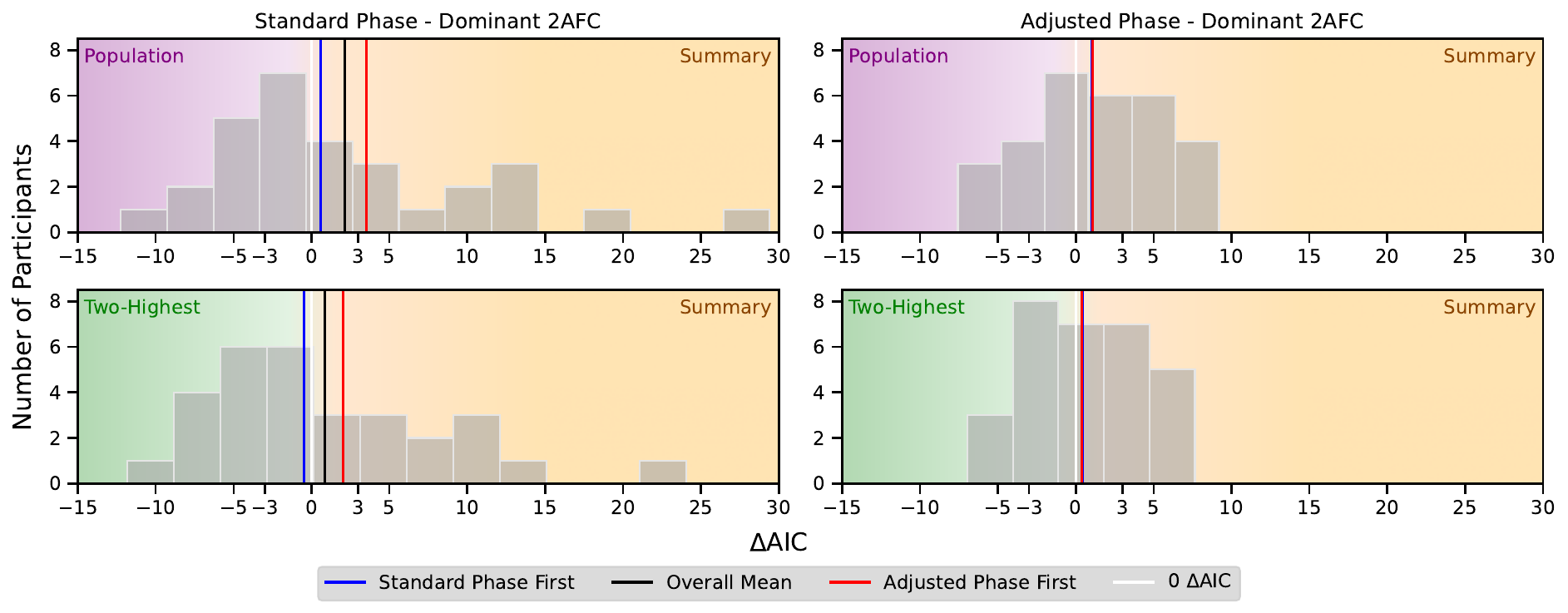}
\caption{$\Delta$AIC of model comparisons of dominant 2AFC trials in Experiment 1. The model with the lowest average AIC (i.e., best fit) was the summary model. For each participant, $\Delta$AIC was computed between the best-fitting summary model and either the two-highest or population model. Positive $\Delta$AIC indicates evidence for the summary model (yellow); negative $\Delta$AIC is evidence for the other model (green or purple). Black lines reflect overall mean $\Delta$AIC across all participants; blue lines indicate mean $\Delta$AIC of participants who underwent the standard phase first; red lines for those who underwent the adjusted phase first; and white lines show $\Delta$AIC $=0$, meaning equal model evidence. The gray bars in the background are histograms of $\Delta$AIC with y-axes of participant count.}
\label{fig: exp1_aic_dominant2AFC}
\end{figure}

Table~\ref{tab:aic_exp1} and Figure~\ref{fig: exp1_aic_dominant2AFC} show AIC differences ($\Delta$AIC) for all three models relative to the best-fitting model (i.e., lowest AIC), the summary model. In the standard phase, the summary model was 1.53 times more likely than the two-highest model (0.85 average $\Delta$AIC) and 2.93 times more likely than the population model (2.15 average $\Delta$AIC; black lines in Figure~\ref{fig: exp1_aic_dominant2AFC}, left panels). However, these AIC differences do not meet a widely held criterion of $\geq$ 3 $\Delta$AIC (4.48 times more likely), suggesting that the data provide no strong preference between the models.

\begin{table}[h!]
\centering
\footnotesize
\caption{$\Delta$AIC scores on dominant 2AFC trials for Experiment 1, where 0 is set to the best-fitting model of a particular phase and order.}
\begin{tabular}{l|ccc|ccc}
\textbf{} 
& \multicolumn{3}{c|}{\textbf{Standard Phase (Dominant 2AFC)}} 
& \multicolumn{3}{c}{\textbf{Adjusted Phase (Dominant 2AFC)}} \\
& \textbf{Overall} & \textbf{Standard First} & \textbf{Adjusted First} 
& \textbf{Overall} & \textbf{Standard First} & \textbf{Adjusted First} \\
Summary       & 0& 0.48 & 0 & 0 & 0 & 0 \\
Two-Highest   & 0.85 & 0 & 2.02 & 0.4 & 0.43& 0.37\\
Population    & 2.15 & 1.07& 3.52 & 1.05 & 1.04& 1.06 \\
Best Model        & Summary       & Two-Highest   & Summary       & Summary       & Summary       & Summary       \\
Second Best & Two-Highest   & Summary       & Two-Highest   & Two-Highest   & Two-Highest   & Two-Highest   \\
\end{tabular}
\label{tab:aic_exp1}
\end{table}

Since phase order was counterbalanced across participants, either the standard or the adjusted phase was encountered first. It is possible that completing the standard phase \textit{after} the adjusted phase influenced model fits on the standard phase. For example, participants who began with the adjusted phase may have adopted a broader attention strategy (see Figure 1), since nondominant trials required considering the full color distribution because the dominant color was absent from the 2AFC options. This strategy could carry over into the standard phase, potentially weakening evidence for the summary model there. To test this, we split the data by phase order. For participants who began with the standard phase, all models still had minimal AIC differences on the standard phase ($\leq1.07$ average $\Delta$AIC; blue lines in Figure~\ref{fig: exp1_aic_dominant2AFC}, left panels). For participants who encountered the standard phase \textit{after} the adjusted phase, the summary model had, in fact, a stronger fit than the population model (3.52 average $\Delta$AIC), though not compared to the two-highest model (2.02 average $\Delta$AIC; red lines Figure~\ref{fig: exp1_aic_dominant2AFC}, left panels). These results indicate that encountering the adjusted phase first did not diminish evidence for the summary model on the standard phase.

In the adjusted phase, differences in model evidence were even less pronounced compared to those in the standard phase, regardless of phase order ($\leq 1.06$ average $\Delta$AIC for all model comparisons). This indicates that the inclusion of nondominant 2AFC trials decreased summary model evidence for the same participants, thereby increasing evidence for the other models, relative to when they were absent (in the standard phase). Thus, prior evidence for the summary model may not reflect a genuine representational capacity limit in the perceptual decision stage.

\section{Experiment 2}
Although Experiment 1 suggests that prior evidence favoring the summary model may reflect resource-rational adaptations to task demands, it does not provide strong support for any particular model, leaving open the question of whether the full distribution of hypotheses is represented at the perceptual decision-stage. To more decisively test between models, Experiment 2 was designed to amplify the predicted accuracy difference between the population and summary models on dominant 2AFC trials by altering the distribution of colors in the stimulus.


\subsection{Method}
\subsubsection{Participants}
30 undergraduates from UNH (who did not participate in Experiment 1) participated for course credit and provided informed consent. All participants had normal or corrected-to-normal visual acuity.

\subsubsection{Stimuli, Design, and Procedure}
\begin{figure}[!b]
\centering
\includegraphics[width=0.65\textwidth]{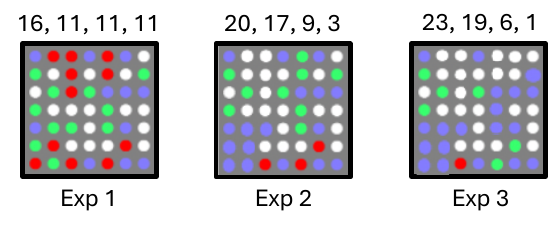}
\caption{Example stimuli with color frequency distribution of Experiment 1 (e.g., 16 white, 11 blue, 11 green, 11 red), Experiment 2 (e.g., 20 white, 17 blue, 9 green, 3 red), and Experiment 3 (e.g., 23 white, 19 blue, 6 green, 1 red.}
\label{fig:color_distribution}
\end{figure}

\shortciteA{yr} mathematically derived the expected accuracy difference between the population and summary models on a dominant 2AFC trial as the expression: $1/3  \big[ P(\text{dominant color is 2nd highest}) -  P(\text{dominant color is lowest})\big]$, where the probability terms reflect the dominant color having a certain rank in the sensory representation. Because these probability terms were not approximated, a numerical estimate of the accuracy difference was not computed. If these terms could be approximated from features of the experiment, one could change the experiment to increase the accuracy difference. To do so, we assumed that the \textit{frequency distribution of the colors in the stimulus} approximates the probability of the dominant color's rank. As an example, in Experiment 1, the color distribution of circles is 16, 11, 11, and 11; thus, the probability that the dominant color holds the highest activation in the sensory representation would be 16/49. Probabilities of other ranks are computed with rank-order statistics (Supplemental Information). To identify a distribution of colors that yields a high accuracy difference between the population and summary models, we iterated over all possible distributions for 49 circles and selected the distribution of 20, 17, 9, and 3 (Figure~\ref{fig:color_distribution}). While this particular distribution does not yield the largest difference possible, it prevents a ceiling effect in participants' accuracy that could arise when relative color frequencies are too extreme and easily distinguishable (which is the case for many distributions with larger accuracy difference than the one chosen here).

The design and procedure were identical to those used with UNH participants in Experiment 1, except for the following difference. Note that the imbalance in nondominant color frequencies (17, 9, 3) creates variability in the difficulty of a 2AFC trial. For example, distinguishing between 20 vs. 3 is easier than 20 vs. 17. If, by chance, a participant encounters a higher proportion of easier 2AFC trials, their 2AFC accuracy will be high while their 4AFC accuracy remains unchanged. This means that the population model will fit better for participants who encountered difficult 2AFC trials more frequently. To control for this, each participant encountered an equal distribution of pairings of color ranks for both dominant and nondominant 2AFC trials.

\subsubsection{Modeling Framework}
To account for the unbalanced color distribution, we estimated three parameters: $\mu_{p,1}, \mu_{p,2}, \mu_{p,3}$ for the dominant, second-most dominant, and third-most dominant colors, respectively, with the least dominant color set to a mean value of 0. To ensure that estimates respected the rank order, a constraint of $\mu_{p,1} > \mu_{p,2} > \mu_{p,3} > 0$ was specified in the NIMBLE package (Supplemental Materials). 

\subsection{Results}
Following the same trial exclusion criteria of Experiment 1 (reaction time below 50 ms or above 4 standard deviations from mean), an average of 14.6 trials (1.27\%) were dropped from the 1152 non-practice trials.

\begin{figure}[H]
\centering
\begin{subfigure}{\textwidth}
    \centering
    \includegraphics[width=\textwidth]{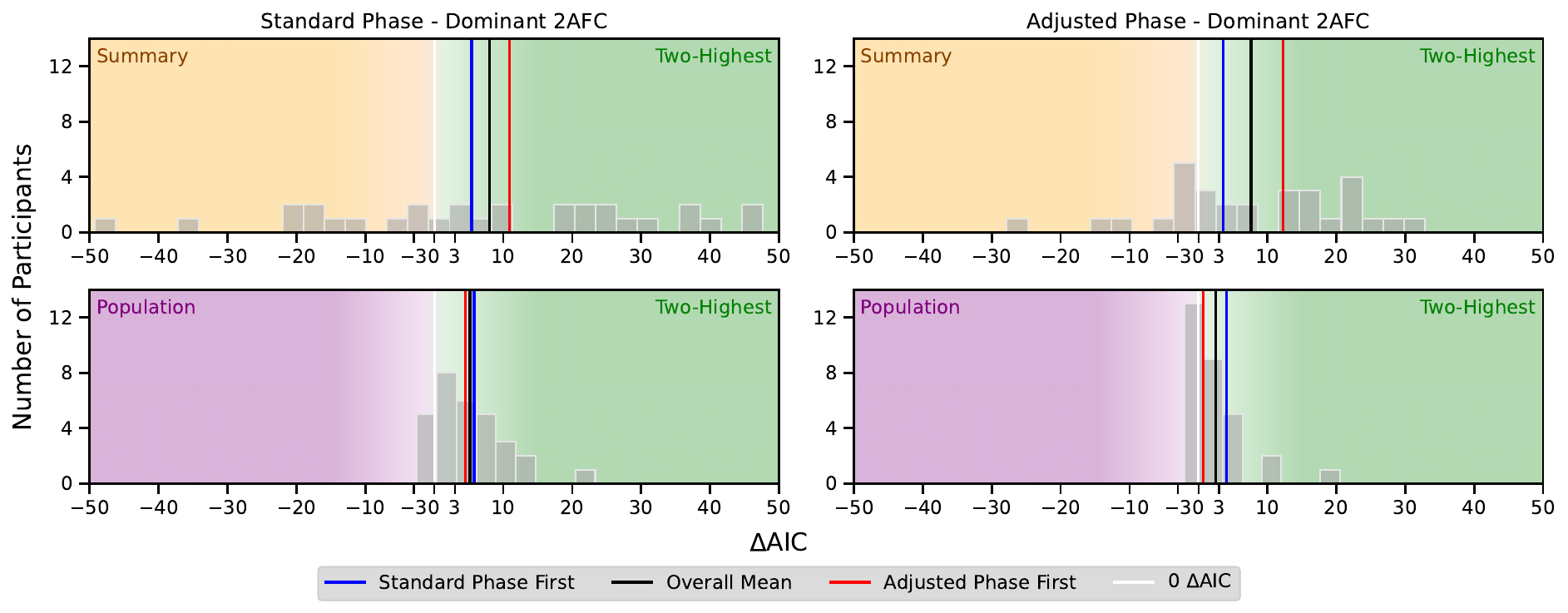}
    \caption{Dominant 2AFC trials}
    \label{fig: exp2_aic_dominant2AFC}
\end{subfigure}
\vspace{1em} 
\begin{subfigure}{0.6\textwidth}
    \centering
    \includegraphics[width=\textwidth]{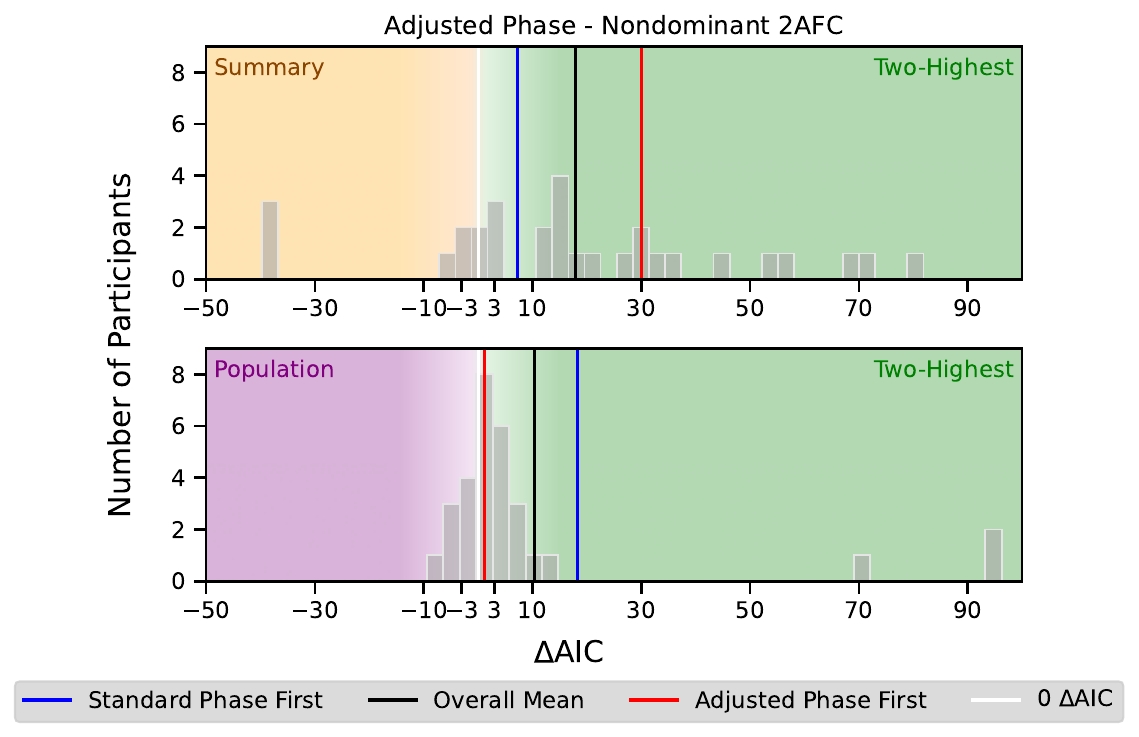}
    \caption{Nondominant 2AFC trials}
    \label{fig: exp2_aic_nondominant2AFC}
\end{subfigure}
\caption{$\Delta$AIC of model comparisons in Experiment 2 for (a) dominant 2AFC trials in each phase and (b) nondominant 2AFC trials in the adjusted phase. Note that the x-axis range is much larger in (b) compared to (a). The model with the lowest average AIC was the two-highest model. For each participant, $\Delta$AIC was computed between the best-fitting two-highest model and either the summary or population model. Positive $\Delta$AIC indicates evidence for the two-highest model (green); negative $\Delta$AIC is evidence for the other model (yellow or purple). Black lines reflect overall mean $\Delta$AIC across all participants; blue lines indicate mean $\Delta$AIC of participants who underwent the standard phase first; red lines for those who underwent the adjusted phase first; and white lines show $\Delta$AIC $=0$, meaning equal model evidence. The gray bars form histograms of $\Delta$AIC with y-axes of participant count.}
\label{fig: exp2_combined}
\end{figure}

For dominant 2AFC trials of the standard phase (Figure~\ref{fig: exp2_aic_dominant2AFC}, left panels), the two-highest model was 53.8 times more likely than the summary model (7.97 average $\Delta$AIC) and 13.3 times more likely than the population model (5.17 average $\Delta$AIC). This advantage held largely regardless of phase order (Table~\ref{tab:aic_exp2_dominant}). For dominant 2AFC trials of the adjusted phase (Figure~\ref{fig: exp2_aic_dominant2AFC}, right panels), the two-highest model was more likely than the summary model (45.4 times more likely; 7.63 average $\Delta$AIC); this was true regardless of phase order. However, the two-highest model was now comparable to or more probable than the population model, depending on phase order. When the adjusted phase preceded the standard phase, there was no evidence for the two-highest model over the population model (0.65 average $\Delta$AIC), but when the adjusted phase followed the standard phase, the two-highest model was more probable than the population model (4.10 average $\Delta$AIC). 

\begin{table}[h!]
\centering
\footnotesize
\caption{$\Delta$AIC scores on dominant 2AFC trials for Experiment 2, where 0 is set to the two-highest model, the best-fitting model within each phase and order.}
\begin{tabular}{l|ccc|ccc}
\textbf{} 
& \multicolumn{3}{c|}{\textbf{Standard Phase (Dominant 2AFC)}} 
& \multicolumn{3}{c}{\textbf{Adjusted Phase (Dominant 2AFC)}} \\
& \textbf{Overall} & \textbf{Standard First} & \textbf{Adjusted First} 
& \textbf{Overall} & \textbf{Standard First} & \textbf{Adjusted First} \\
Summary       & 7.97& 5.42& 10.88& 7.63& 3.61& 12.22\\
Two-Highest   & 0 & 0 & 0           & 0     & 0     & 0  \\
Population    & 5.17& 5.78& 4.48& 2.49& 4.10& 0.65\\
Best Model        & Two-Highest   & Two-Highest   & Two-Highest   & Two-Highest   & Two-Highest   & Two-Highest   \\
Second Best & Population    & Population    & Summary       & Population    & Population    & Population    \\
\end{tabular}
\label{tab:aic_exp2_dominant}
\end{table}

A similar pattern emerges when fitting nondominant 2AFC data of the adjusted phase (which, unlike in Experiment 1, now had ground-truth answers due to the unbalanced color frequency distribution). On these trials, the two-highest model overwhelmingly outfit the summary model (7644 times more probable; 17.88 average $\Delta$AIC) and was either comparable to or much more probable than the population model, depending on phase order (Figure~\ref{fig: exp2_aic_nondominant2AFC}). When the adjusted phase preceded the standard phase, there was no evidence for the two-highest model over the population model (1.21 average $\Delta$AIC), but when the adjusted phase followed the standard phase, the two-highest model was more probable than the population model (18.26 average $\Delta$AIC). 

\begin{table}[h]
\centering
\footnotesize
\caption{$\Delta$AIC scores on nondominant 2AFC trials for Experiment 2, where 0 is set to the two-highest model, the best-fitting model within each order.}
\begin{tabular}{l|ccc}
\textbf{} 
& \multicolumn{3}{c}{\textbf{Adjusted Phase (Nondominant 2AFC)}} \\
& \textbf{Overall} & \textbf{Standard First} & \textbf{Adjusted First} \\
Summary       & 17.88 & 7.21  & 30.08 \\
Two-Highest   & 0         & 0     & 0 \\
Population    & 10.30 & 18.26 & 1.21 \\
Best Model        & Two-Highest   & Two-Highest   & Two-Highest   \\
Second Best & Population    & Summary       & Population    \\
\end{tabular}
\label{tab:aic_exp2_nondominant}
\end{table}

In sum, for dominant 2AFC trials of the standard phase, the two-highest model consistently outperformed both the summary and population models. For both types of 2AFC trials in the adjusted phase, it continued to have an advantage over the summary model, but its performance relative to the population model varied. When participants encountered the adjusted phase first, the population model performed similarly as the two-highest model. When the standard phase preceded the adjusted phase, the two-highest model either outperformed or performed similarly as the population model. This suggests that a resource-rational strategy from the standard phase may have carried over into the adjusted phase, confounding evidence for the two-highest model.

These results are consistent with the fact that retaining the two most dominant colors in the perceptual decision-stage representation is sufficient for the model to achieve high accuracy for our experiment. Indeed, in the standard phase, where the dominant color is always present among response options, representing the two most dominant colors reliably leads to the correct answer. In the adjusted phase as well, this strategy is effective, failing only in the rare case where the two least dominant colors are paired together as response options. Table~\ref{tab:trial_distribution_exp2} illustrates this case, with the highlighted row indicating where the two-highest model yields an inaccurate response.

\begin{table}[H]
\centering
\footnotesize
\caption{Example accuracies for trials in the adjusted phase of Experiment 2 using a representation of the two most dominant colors at the perceptual decision stage. The example stimulus has a color frequency distribution of 20, 17, 9, and 3. The highlighted row indicates the trial type where the two-highest model yields an inaccurate response.}
\begin{minipage}{0.15\textwidth}
    \centering
    \includegraphics[width=\textwidth]{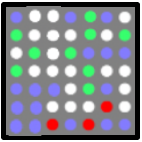} 
\end{minipage}%
\hfill
\begin{minipage}{0.8\textwidth}
    \centering
    \begin{tabular}{cccc}
        \textbf{Trial Type} & \textbf{Options (as ranks)} & \textbf{Correct Answer} & \textbf{Accuracy} \\ 
        4AFC                  & 1, 2, 3, 4 & 1 (white) & Correct \\ 
        Dominant 2AFC         & 1, 2       & 1 (white) & Correct \\ 
        Dominant 2AFC         & 1, 3       & 1 (white) & Correct \\ 
        Dominant 2AFC         & 1, 4       & 1 (white) & Correct \\ 
        Nondominant 2AFC      & 2, 3       & 2 (blue) & Correct \\ 
        Nondominant 2AFC      & 2, 4       & 2 (blue) & Correct \\ 
        \rowcolor[HTML]{FFFF99}
        Nondominant 2AFC      & 3, 4       & 3 (green) & Incorrect \\ 
    \end{tabular}
\end{minipage}
\label{tab:trial_distribution_exp2}
\end{table}

\section{Experiment 3}
Representing the two most dominant colors was a near-optimal resource-rational strategy in Experiment 2. To test whether participants can represent the full distribution at the decision stage, Experiment 3 modified the trial distribution of the experiment so that representing any subset of colors short of all four would result in chance or below-chance accuracy (except a few subsets). We also amplified the predicted accuracy difference between the two-highest and population models by introducing a new stimulus color distribution within this modified trial distribution. Participants underwent only the adjusted phase; the standard phase was removed to prevent potential carry-over of resource-rational strategies, as discussed in Results of Experiment 2.


\subsection{Method}
\subsubsection{Participants}
30 undergraduates from UNH (who did not participate in prior experiments) participated for course credit and provided informed consent. All participants had normal or corrected-to-normal visual acuity.

\subsubsection{Stimuli, Design, and Procedure}
In prior experiments, adjusted-phase trials were evenly divided among 4AFC, dominant 2AFC, and nondominant 2AFC trials. In Experiment 3, we increased the proportion of nondominant 2AFC trials featuring the two least dominant colors. This trial type is the only trial in which the two-highest and population models diverge: the two-highest model fails, as it represents only the top two colors, whereas the population model succeeds. However, setting the entire experiment to 2AFC trials with the two least dominant colors might lead participants to adopt a strategy of focusing on the two least colors. Given this, we created a new trial distribution that assigns one-fourth of trials as 4AFC, removes dominant 2AFC trials, and distributes the remainder across nondominant 2AFC subtypes. An example with eight total trials is provided in Table~\ref{tab:trial_distribution_exp3}.

\begin{table}[h!]
\centering
\footnotesize
\caption{Trial distribution for Experiment 3 if the experiment consisted of eight total trials. All 2AFC trials are nondominant; 4AFC trials are included to estimate sensory representations. Colors are intended to distinguish trial subtypes (e.g., nondominant 2AFC with rank 2 and 3 versus with rank 2 and 4).}
\begin{tabular}{cccc}
\textbf{Trial Type}      & \textbf{Options (as ranks)}       \\ 
Nondominant 2AFC         & \textcolor[HTML]{00AA00}{2, 3}                     \\ 
Nondominant 2AFC         & \textcolor[HTML]{FF8C00}{2, 4}                     \\ 
Nondominant 2AFC         & \textcolor[HTML]{0000FF}{3, 4}                     \\ 
Nondominant 2AFC         & \textcolor[HTML]{0000FF}{3, 4}                     \\ 
Nondominant 2AFC         & \textcolor[HTML]{0000FF}{3, 4}                     \\ 
Nondominant 2AFC         & \textcolor[HTML]{0000FF}{3, 4}                     \\
4AFC                     & \textcolor[HTML]{FF0000}{1, 2, 3, 4}               \\ 
4AFC                     & \textcolor[HTML]{FF0000}{1, 2, 3, 4}               \\ 
\end{tabular}
\label{tab:trial_distribution_exp3}
\end{table}

This revised trial distribution ensures that attending to or retaining any subset of colors other than all four would result in chance or below-chance accuracy, except for unlikely combinations of two colors. For instance, 50\% accuracy is achieved by representing only the two most dominant colors (accurate for the four red, green, and orange rows, but not the other four blue rows in Table~\ref{tab:trial_distribution_exp3}) or the two least dominant colors (accurate for the four blue rows, but not the other four rows in Table~\ref{tab:trial_distribution_exp3}). Other pairings---such as the dominant and least dominant (accurate for the two red rows), or the second most and least dominant colors (accurate for the two green and orange rows)---yield only 25\%. Following similar logic, representing a single color never exceeds 50\% accuracy, regardless of the color's rank. Accuracy can reach 75\%, but only by representing highly specific pairs (either the second and third most dominant colors, or the dominant and third most dominant). Sustaining such selective combinations across the experiment is implausible.

Using this revised trial distribution, we selected a new stimulus color distribution of 23, 19, 6, and 1 (Figure~\ref{fig:color_distribution}) by iterating over all possible color distributions (as in Experiment 2), but this time increasing the accuracy difference between the \textit{two-highest} and population models on nondominant 2AFC trials of the modified trial distribution (Supplemental Information). Each block contained 40 trials. Participants completed 15 blocks plus 1 practice block, totaling 640 trials. Breaks were also shortened: starting with the first non-practice block, each break lasted 10 seconds (previously 15 seconds), and a 30-second break was given every 8 blocks (previously 1 minute).

\subsection{Results}
\begin{figure}[!b]
\centering
\includegraphics[width=0.5\textwidth]{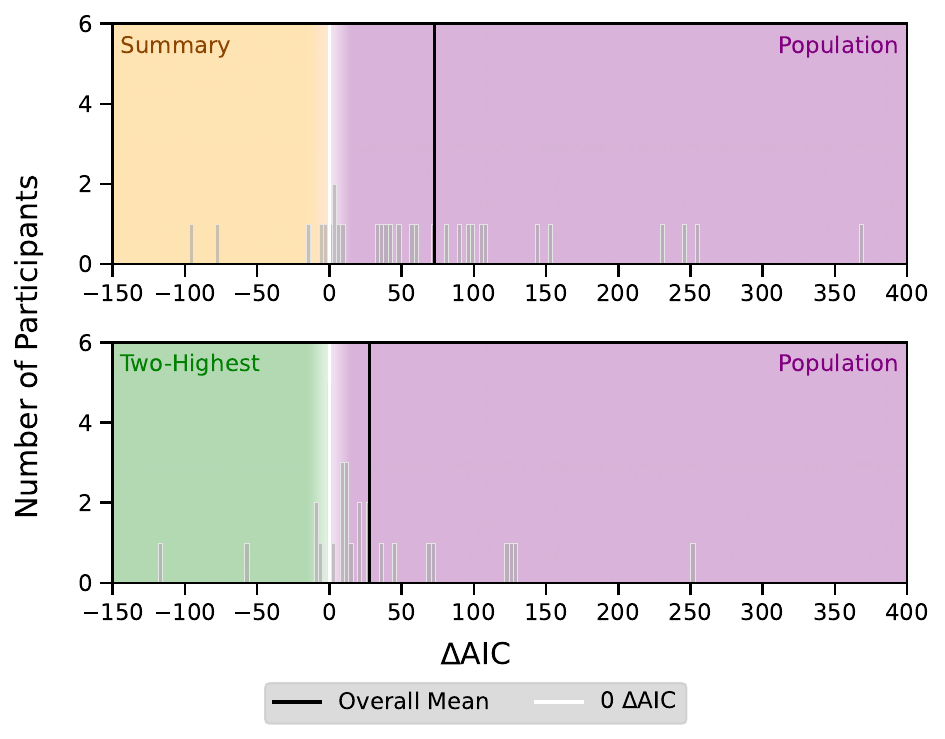}
\caption{$\Delta$AIC of models fit to the specific variants of nondominant 2AFC trials of Experiment 3. The model with the lowest average AIC was the population model. For each participant, $\Delta$AIC was computed between the best-fitting population model and either the summary or two-highest model. Positive $\Delta$AIC indicates evidence for the population model (purple); negative $\Delta$AIC is evidence for the other model (yellow or green). Note that the x-axis range is much larger than in previous plots, and there are no blue or red lines as there is only one experimental phase. White lines show $\Delta$AIC $=0$, meaning equal model evidence. The gray bars in the background are histograms of $\Delta$AIC with y-axes of participant count.}
\label{fig:exp3_aic_nondominant2AFC}
\end{figure}

Using the same reaction-time criteria as before (reaction time below 50 ms or above 4 standard deviations from mean), an average of 5.23 trials (0.87\%) were dropped from the 600 experimental trials. The population model provided the best fit to nondominant 2AFC data, followed by the two-highest model and then the summary model (Table~\ref{tab:aic_exp3}). The population model was 9.8 $\cdot 10^{5}$ times more likely than the two-highest model (27.6 average $\Delta$AIC) and 6.5 $\cdot 10^{15}$ times more likely than the summary model (72.83 average $\Delta$AIC). See Figure~\ref{fig:exp3_aic_nondominant2AFC} for a visual.

\begin{table}[H]
\centering
\footnotesize
\caption{$\Delta$AIC scores on nondominant 2AFC trials for Experiment 3, where 0 is set to the population model, the best-fitting model.}
\begin{tabular}{l|c}
\textbf{} & \textbf{Modified Trial Distribution} \\
Summary       & 72.83 \\
Two-Highest   & 27.60 \\
Population    & 0 \\
Best Model    & Population \\
Second Best   & Two-Highest \\
\end{tabular}
\label{tab:aic_exp3}
\end{table}

\section{Discussion}

In this study, we tested whether complex representations are maintained and used at the perceptual decision stage. Our hypothesis was that seemingly simple decision-stage representations (e.g., \shortciteNP{yr}; see \shortciteNP{rahnev2018suboptimality} for a review) arise not from limits in perception, but participants' resource-rational adaptations to task demands. To fairly test for complex representations, we carefully controlled a decision task so that success increasingly required the use of a complex representation, such as a full distribution, while limiting the efficacy of resource-rational strategies. 

In Experiment 1, we interleaved trials where the dominant color was excluded from response options, forcing participants to consider information beyond the most dominant color. This initial design produced comparable fits between representationally simple and complex decision models. In Experiment 2, we modified the stimulus to sharpen the divergence between model predictions. This reduced evidence for the summary model and increased evidence for the two-highest and population models. In Experiment 3, we modified the stimulus and the experiment's trial distribution to further mitigate resource-rational strategies, producing clear evidence for the population model. This systematic progression of evidence toward the population model as task demands grew stringent suggests that participants' decisions relied on complex representations when necessary, and simpler ones otherwise. Therefore, (1) suboptimal performance reported in previous studies may reflect participants' resource-rationality strategies, rather than a smaller capacity of decision-making circuits relative to sensory circuits \shortcite{yr}; and (2) to our knowledge, we demonstrate the closest approximation of direct evidence for complex representations in perception. 

Our findings add an epistemic constraint to the debate around probabilistic representations: researchers should consider whether their tasks permit high performance through resource-rational strategies that could mask complex representations \shortcite{lieder2020resource}. The allocation of attention is widely considered a cost-benefit tradeoff between limited attentional resources and task performance (e.g., \shortciteNP{butko2008pomdp, van2018resource}). Since task success did not require full stimulus attention in \shortciteA{yr} and the standard phase of Experiment 1, participants' suboptimal performance may reflect a trade-off in consistently identifying the dominant color throughout our long, repetitive experiment. Likewise, since success required greater stimulus attention in Experiments 2 and 3, optimal-like performance may reflect participants' willingness to incur higher attentional costs over a shorter experiment. Supporting this view, \shortciteA{lindig2022bayes} showed that under time pressure participants relied on a single, diagnostic feature when classifying simple objects composed of three features, but integrated multiple features with more time. 

As \shortciteA{rahnev2022mystery} argued, we demonstrate that progress can be made when framing the debate in terms of representational complexity. Although our findings do not necessarily show evidence of \textit{strong} probabilistic representations---literal probability distributions defined by Kolmogorov axioms---since all tested decision models use unitless values \shortcite{rahnev2021perception}, they go beyond a \textit{weak} definition---where one observes mere variation in participants' response frequencies---because complex representations better explain the data than simpler ones. Of course, this argument is valid insofar as models based on complex representations consistently outperform simpler ones across a range of perceptual phenomena.

Future work is needed to determine whether complex representations also emerge in more naturalistic tasks and environments, where perception operates without many constraints. Like many forced-choice tasks, participants in our experiment knew the goal (select more dominant color), the answer format (colors), and the exact set of possibilities (four predefined colors). For this reason, rather than focusing solely on the empirical evidence for complex representations, we emphasize that resource rationality is both an epistemic constraint for studies of representations and essential to a complete theory of perception.

\section{Acknowledgments}
We thank Nick Ichien, Caroline Curry, Julia Matin, and Katrin Aspelund.


\bibliography{citations}
\bibliographystyle{apacite}    

\section{Supplemental Information}

\subsection{No qualitative differences between participant groups in Experiment 1}
To confirm that the two undergraduate participant groups in Experiment 1 do not qualitatively differ, we separately analyzed accuracy and reaction time. For accuracy, we ran a three-way mixed ANOVA between sample (Williams College/University of New Hampshire), phase (standard/adjusted), and trial type (4AFC/dominant 2AFC). We used only two levels in the trial type factor (the two trial types shared by both phases), as the nondominant 2AFC trials in the adjusted phase lack correct answers in Experiment 1. This three-way ANOVA revealed no three-way interaction ($F(1,28) = 0.28, p = 0.60$), no two-way interaction between sample and phase ($F(1,28) = 0.048, p = 0.83$) and between sample and trial type ($F(1,28) = 0.005, p = 0.95$), and no main effect of sample ($F(1,28) = 1.57, p = 0.22$).

Since reaction time is independent of whether a trial has a correct answer, and participants do not know that some trials lack a correct answer, our analyses of reaction time included dominant 2AFC trials. This inclusion resulted in an unequal factorial design, where the standard phase does not, but the adjusted phase does, have a level in the trial type factor for dominant 2AFC trials. To avoid complexities with the unbalanced design, we split the data by phase to run two separate ANOVAs: (1) a two-way repeated-measures (RM) ANOVA between sample and trial type for only standard phase data, and (3) the same ANOVA for only adjusted phase data. For both phases, there was no interaction between sample and trial type (standard phase data: $F(1,28) = 0.42, p = 0.84$; adjusted phase data: $F(1,28) = 0.33, p = 0.57$) and no main effect of sample (standard phase data: $F(1,28) = 1.12, p = 0.30$; adjusted phase data: $F(1,28) = 1.30, p = 0.26$).

Prior to these ANOVAs, outlier trials were dropped for each participant. Excluding the practice block, a trial was dropped if reaction time was faster than 0.05 seconds or more than 4 standard deviations from the participant's mean reaction time. The average number of dropped trials does not differ by sample ($t(28) = 0.19, p = 0.90$). An average of 12 (1\%) and 11.58 (1\%) trials out of 1152 total were dropped for the Williams and UNH students, respectively. As these analyses paint no significant difference across samples, we combine them in the main report.

\subsection{Modeling diagnostics}
Using three common diagnostics in Bayesian modeling, we ensured that each participant's posterior samples (obtained using slice sampling) are representative of their posterior distribution and sufficiently accounts for the data from which they was derived. For each posterior sample, we computed the Gelman-Rubin convergence diagnostic, its effective sample size, and a posterior predictive check. 

The Gelman-Rubin convergence diagnostic measures whether an MCMC simulation of a parameter has converged onto a stable set of samples by calculating the similarity of variance across multiple simulations with different initializations. A lower Gelman-Rubin diagnostic (minimum 1) indicates higher similarity across simulations. For each participant, three simulations were run. In Experiment 1, $\mu$ was initialized as a random normal draw with mean 1 and standard deviation 0.5. All participants' simulations of $\mu$ in Experiment 1 had a Gelman-Rubin diagnostic between 1 and 1.05 (a standard acceptable ceiling). In Experiment 2, $\mu$, $\mu_{N1}$ and $\mu_{N2}$ were initialized as a random normal draw with mean 1, 0.75, and 0.5 and standard deviation 0.2, respectively. Since Experiment 2 contains multiple parameters, we calculated the point-scale reduction factor (PSRF) of each participant, a summary score of the Gelman-Rubin diagnostics of the individual parameters. Participants' simulations, except for seven, had a PSRF between 1 and 1.05. For the remaining seven, a PSRF less than 1.05 was achieved by re-running simulations with a larger simulation of 1,050,000 samples and 50,000 burn-in and with new initializations of $\mu$, $\mu_{N1}$, and $\mu_{N2}$ centered at 3, 2, and 1 (determined post-hoc by calculating the average posterior means of participants with an acceptable PSRF).

After ensuring an acceptable Gelman-Rubin diagnostic for all participants, we calculated the effective sample size (ESS) of each posterior sample. Since a sample in a MCMC simulation is dependent on the previous sample, the simulation is not guaranteed to be representative of the estimated posterior distribution. ESS captures the representativeness of a posterior sample by measuring the extent of auto-correlation of the sample/simulation over time, where the lower the auto-correlation, the closer the simulation is to a simple random sample and the higher the ESS. In Experiment 1, the average ESS of all samples of $\mu$ across both phases was 72430.53 (SD = 7014.58). In Experiment 2, the average ESS of all samples of $\mu$, $\mu_{N1}$, and $\mu_{N2}$ were 6054.46 (SD = 7384.85), 6064.34 (SD = 7359.95), and 9453.22 (SD = 12464.27) respectively. 

Finally, we performed a posterior predictive check of each simulation by determining whether the simulation can sufficiently recreate the 4AFC accuracy from which it was estimated. In Experiment 1, each participant's correct accuracy on 4AFC trials (separately per phase) was predicted 100 times from their posterior mean of $\mu$. Predicted accuracy was a proportion out of the same number of 4AFC trials that were not dropped (see outlier criteria in Results sections) for that participant's phase. Across participants, the average error of each prediction to the true accuracy was $0.09\%$. Similarly, in Experiment 2, each participant's accuracy was predicted 100 times from their posterior means of $\mu$, $\mu_{N1}$ and $\mu_{N2}$. Similarly, across participants, the average error of each prediction was $0.06\%$.

\subsection{Amplifying accuracy differences by changing the color frequency distribution}
In Experiment 2, we changed the color frequency distribution to 20, 17, 9, and 3 from 16, 11, 11, and 11 in Experiment 1. In Experiment 3, we again changed the distribution to 23, 19, 6, and 1. As outlined in the main text, we predicted that changing the color frequency distribution would influence the difference in accuracy predictions of the population and summary models, thereby increasing the strength of evidence for one model over another. To demonstrate this, we start with a summary of Yeon and Rahnev's (2020) mathematical derivation of predicted model accuracies on dominant 2AFC trials.

\subsubsection{Derivation of model accuracies on dominant 2AFC trials}
\shortciteA{yr} begin with the observation that the probability of being correct on a dominant 2AFC trial depends on the ``ranking'' of the dominant color relative to the nondominant colors (i.e., the relative magnitude of the dominant color activation when compared to the activations of the other colors). Mathematically, this is equivalent to:
\begin{align*}
    P(y=1) &= P(y=1|D_1)P(D_1) + P(y=1|D_2)P(D_2) + P(y=1|D_3)P(D_3) + P(y=1|D_4)P(D_4)
\end{align*}
where $P(D_i)$ is the probability that the dominant color has the $i$th highest ranking.

The value of the conditional coefficients of this equation (e.g., $P(y=1|D_i)$) depends on whether one chooses a population or summary model decision-making strategy. If $i=1$ (i.e., dominant color has the maximum activation), both models are always correct: $P(y=1|D_1) = 1$. The models differ in the other cases.

When $i=2$, the population model will be correct, so long as the nondominant alternative is not the highest activation. Since the experiment randomizes which of the three nondominant colors is presented in the dominant 2AFC trial, the likelihood that the experiment chooses the nondominant color with the maximum activation is $1/3$. Thus, the population model will be correct $2/3$ of the time: $P(y=1|D_2) = 2/3$. Similarly, when $i=3$, the population model is correct when the nondominant alternative has the lowest activation. The chances of this happening are $1/3$, so the population model will be correct $1/3$ of the time. Finally, when $i=4$, the nondominant alternative always has higher activation than the dominant color, so the population model is always incorrect.

Like the population model, when $i=2$, the summary model is correct so long as the nondominant alternative is not the highest activation, but only 50\% of the time. This is because neither the dominant color or the nondominant alternative are the maximum, so the summary model selects one of the options at random. Since, as mentioned earlier, $2/3$ is the probability the nondominant color is not the highest activation, the summary model is correct $2/3 \cdot 1/2 = 1/3$ of the time: $P(y=1|D_2) = 1/3$. This reasoning is, interestingly, the exact same as when $i=3$ or $i=4$.

By plugging these values into the coefficients above, and then subtracting the resultant population and summary models' probabilities, we obtain: $P(y=1)_{pop} - P(y=1)_{sum} = \frac{1}{3}P(D_2) - \frac{1}{3}P(D_4)$. But since $P(D_2)$ is greater than $P(D_4)$ (i.e., it's more likely that the dominant color has a higher activation than not), the overall difference must be positive. Thus, the population model should be more accurate than the summary model on dominant 2AFC trials (\emph{independent} of participant model fitting; the models can have similar accuracies once fit onto a dataset). 

\subsubsection{Completing the derivation using the color frequency distribution}
We complete Yeon and Rahnev's (2020) derivations by providing numeric estimates for the accuracies of the population and summary models. We do so by proposing a way to calculate the probability that the dominant color is a given rank $P(D_i)$ from the color frequency distribution of the stimulus. Assume that the dominant color has 16 circles in the array, and the nondominant colors have 11 each, as in Experiment 1. There are 49 circles in total, and therefore a dominant color proportion of $16/49$ in the array. We can think of this proportion as an approximation of perceptual salience, or how strongly the dominant color appears relative to the other colors. 

Computing the probability that the dominant color has the highest activation is trivial. Calculating its probability for other ranks (e.g., $P(D_3)$) is more difficult and involves order statistics. This is because when the dominant color is a specific rank, each of the other colors must have a specific rank as well. One particular rank order may differ in probability of occurrence from another, even if the dominant color has the same rank in both. Thus, we need to rewrite Yeon and Rahnev's (2020) original expression for a model's probability of being correct by marginalizing over all rank orders:
\begin{align*}
    P(y=1) &= P(y=1|D_1,N1_2,N2_3,N3_4) \space \cdot \space P(D_1,N1_2,N2_3,N3_4) \\
    &+ P(y=1|D_1,N1_3,N2_2,N3_4) \space \cdot \space P(D_1,N1_3,N2_2,N3_4) \\
    &+ ...
\end{align*}
where $N1, N2, N3$ and $N4$ are the three nondominant colors and the subscripts reflect their relative ranks. Calculating the conditional coefficients of this equation (e.g., $P(y=1|D_1,N1_2,N2_3,N3_4)$) follows the same reasoning as before. Calculating the probability of a particular rank order, such as $P(D_1, N1_2, N2_3, N3_4)$, requires the following probability formula:
\begin{align*}
    P(D_1, N1_2, N2_3, N3_4) &= P(D_1) \times \frac{P(N1_1)}{1-P(D_1)} \\
    &\times \frac{P(N2_1)}{1-P(D_1)-P(N1_1)} \\
    &\times \frac{P(N3_1)}{1-P(D_1)-P(N1_1)-P(N2_1} \\
\end{align*}
where $P(D_1), P(N1_1), P(N2_1), P(N3_1)$ depend on the color frequency distribution, such as 16, 11, 11, and 11:
\begin{align*}
    P(D_1) &= 16/49 \\
    P(N1_1) &= 11/49 \\
    P(N2_1) &= 11/49 \\
    P(N3_1) &= 11/49 \\
\end{align*}
Substituting these values into the equation above gives us the following predicted model accuracies with a color distribution of 16, 11, 11, and 11:
\begin{align*}
    P(y=1)_{pop} &= 0.593 \\
    P(y=1)_{sum} &= 0.551 \\
    P(y=1)_{pop} - P(y=1)_{sum} &= 0.042 \\
\end{align*}

The accuracy difference between the population and the summary models is small (0.0416). To discover a color frequency distribution that yields a larger difference, we iterate over all possible color frequency distributions that sum to the total number of circles of 49. The distribution with the maximum difference is 33, 12, 3, and 1, with an accuracy difference of 0.25. However, we decided against this distribution because the dominant color had many more circles than the nondominant colors (33 $>$ 12, 3, 1), meaning that the dominant color would appear very salient in the stimulus. This would result in 4AFC and nondominant 2AFC trials that would be too easy for participants, causing a ceiling effect in participants' accuracies. To avoid a ceiling effect, we chose a distribution of 20, 17, 9, and 3 for Experiment 2:
\begin{align*}
    P(y=1)_{pop} &= 0.700 \\
    P(y=1)_{sum} &= 0.605 \\
    P(y=1)_{pop} - P(y=1)_{sum} &= 0.095 \\
\end{align*}

\subsubsection{Derivation of model accuracies on nondominant 2AFC trials}
To derive model accuracies on nondominant 2AFC trials, we start with the same marginalization as above. The probability of obtaining a rank order follows the same probability formula as before. Again, the conditional coefficients differ depending on the population or summary model. 

A nondominant 2AFC trial is constructed by presenting two random nondominant colors $NJ$ and $NK$. There are three possible nondominant 2AFC trial types: (1) $J=1, K=2$, (2) $J=1, K=3$, and (3) $J=2, K=3$, where $J$ is the index of the larger nondominant color and $K$ is the index of the smaller. Given a rank order of $D_1, N1_2, N2_3, N3_4$, the population model is correct in all nondominant 2AFC trial types because $NJ$ is always ranked higher than $NK$. In contrast, the summary model is correct 50\% of the time, since the maximum activation is the dominant color $D_1$, so the model will toss a coin when deciding between the two nondominant alternatives. 

Therefore, the conditional probability of being correct on this rank order for the population model is $P(y=1|D_1, N1_2, N2_3, N3_4) = (1 + 1 + 1)/3 = 1$ and for the summary model is $P(y=1|D_1, N1_2, N2_3, N3_4) = (0.5 + 0.5 + 0.5)/3 = 0.5$.

We iteratively calculate the conditional probabilities of being correct on the three possible nondominant 2AFC trials for every rank order. For a color ratio of 16, 11, 11, and 11, we get the following accuracies:
\begin{align*}
    P(y=1)_{pop} &= 0.5 \\
    P(y=1)_{sum} &= 0.5 \\
    P(y=1)_{pop} - P(y=1)_{sum} &= 0 \\
\end{align*}
This matches with intuition. When the nondominant colors have the same number of circles (each 11), the population model cannot determine which nondominant alternative is greater than the other, even though it has access to information about both. So, the population model will select at random, just like the summary model which does not have access to information other than the color with the maximum activation.

When the color ratio is 20, 17, 9, and 3, we should see that the population model performs better than the summary model because it has access to information about nondominant colors that can now be distinguished. Indeed, we obtain the following accuracies:
\begin{align*}
    P(y=1)_{pop} &= 0.751 \\
    P(y=1)_{sum} &= 0.595 \\
    P(y=1)_{pop} - P(y=1)_{sum} &= 0.156 \\
\end{align*}

These results now reveal three motivation for using 20, 17, 9, and 3 as the color frequency distribution. First, we use 20, 17, 9, and 3 to observe a larger strength of evidence for one model over another on dominant 2AFC trials. Second, this distribution makes it possible to distinguish models on nondominant 2AFC trials, as there are an unequal number of circles across the nondominant colors. Third, we may see that evidence for a model is larger on nondominant 2AFC trials than on dominant 2AFC trials because the predicted accuracy difference is larger ($0.156 > 0.0945$).

\subsection{Model Fit and Log-Likelihood Calculations}
We assess the strength of model fit on observed dominant 2AFC data by calculating the AIC for the population and summary models \shortcite{bozdogan1987model}. AIC is defined as $\text{AIC} = 2\log(\mathcal{L}) + 2k$ where $\mathcal{L}$ is the log-likelihood of the model and $k$ is the number of free parameters in the model. As Yeon and Rahnev (2020) note, $k = 0$ for both the population and summary models since they use the same static parameter values from the fitted sensory representation. 

We define the log-likelihood of a model as 
\begin{align*}
  \log(\mathcal{L}) = \sum_{i=1} \log(\bar{L}(y_i))
\end{align*}
where $i$ is a dominant 2AFC trial and $\bar{L}$ is the average likelihood of obtaining the human's accuracy of $y$ on trial $i$, averaged across every $500th$ posterior sample of $\mu$. Specifically, for a given trial $i$ and a given sample of $\mu$ among all $500th$ samples of $\mu$ in the posterior sample (i.e., $\mu_s$ where $s=1,501,1001,\ldots,S$), we calculate a likelihood of obtaining the human's accuracy $y_i$ on sample $s$: $L(y_i | \mu_p, \; \sigma_{D,s}, \; \sigma_{N,s})$. Then, we average the likelihoods across samples $s = 1,501,1001,\ldots,S$:
\begin{align*}
  \bar{L}_i = \frac{1}{S} \sum_{s=1} L(y_i | \mu_p, \; \sigma_{D,s}, \; \sigma_{N,s})
\end{align*}

The likelihood $L(y_i | \mu_p, \; \sigma_{D,s}, \; \sigma_{N,s})$ of a human's accuracy $y$ for a given trial $i$ and a given sample $s$ is computed by simulating 500 events. In each event, activations for all colors are drawn from normal distributions (centered at $\mu$ for the dominant color and 0 for the nondominant colors, with standard deviation of 1). Each model computes an accuracy of 1 or 0 based on the activations of this event. If the model's accuracy matches the human's accuracy, the likelihood is 1, otherwise it is 0. The average likelihood across all 500 events is $L(y_i | \mu_p, \; \sigma_{D,s}, \; \sigma_{N,s})$.

\end{document}